\documentclass{IEEEtran}

\usepackage{amsmath}
\usepackage{amssymb}
\usepackage{graphicx}
\usepackage{xcolor}
\usepackage{latexsym}
\usepackage{cite}
\usepackage{url}

\usepackage{epic,eepic}

\newtheorem{corollary}{Corollary}

\newtheorem{lemma}{Lemma}
\newtheorem{theorem}{Theorem}

\newtheorem{remark}{Remark}

\newcommand\bydef{\triangleq}

\newcommand\Bc   {{\cal B}}

\newcommand\code {{\cal C}}

\newcommand\Nc {{\cal N}}

\newcommand\Tc  {{\cal T}}

\newcommand\Xc   {{\cal X}}
\newcommand\xc  {{\cal X}}

\newcommand\yc  {{\cal Y}}

\newcommand\tie  {{\tt T}}
\newcommand\notie {{\tt N}}

\definecolor{dg}{rgb}{0, 0.5, 0}

\definecolor{dg}{rgb}{0, 0.5, 0}

\begin{document}

\title{The Asymptotic Generalized Poor-Verd\'u Bound Achieves the BSC Error Exponent at Zero Rate}
%{Exponential Tightness of the Generalized Poor-Verd\'u Error Bound at Zero Rate for the BSC}

\author{
\IEEEauthorblockN{
     Ling-Hua Chang\IEEEauthorrefmark{1},
     Po-Ning~Chen\IEEEauthorrefmark{2},
     Fady Alajaji\IEEEauthorrefmark{3}
     and 
     Yunghsiang S.~Han\IEEEauthorrefmark{4}
\thanks{\IEEEauthorrefmark{1}Dept.~Elec.~Eng., Yuan-Ze University, Taiwan, R.O.C.}     
\thanks{\IEEEauthorrefmark{2}Dept.~Elec.~and Comp.~Eng., Nat'l.\ Chiao Tung Univ., Taiwan, R.O.C.}
\thanks{\IEEEauthorrefmark{3}Dept.~Mathematics  \&  Statistics,  Queen's Univ.,  Kingston, ON,  Canada.}
\thanks{\IEEEauthorrefmark{4}School of Elec.~Eng.~\& Intelligentization,  Dongguan Univ.~of Tech., China.}
\thanks{\footnotesize \{iamjaung,\,qponing\}@gmail.com,\,fa@queensu.ca,\,yunghsiangh@gmail.com}
}
}

\maketitle

\begin{abstract} 
The generalized Poor-Verd\'u error lower bound for multihypothesis testing is revisited. 
Its asymptotic
expression is established in closed-form as its tilting parameter grows to infinity. 
It is also shown that the 
asymptotic generalized bound achieves the error exponent (or reliability function) of the
memoryless binary symmetric channel at zero coding rates.
\end{abstract}


\section{Introduction}

A well-known lower bound on the minimum probability of error $P_e$ of multihypothesis testing 
is the so-called Poor-Verd\'{u} bound \cite{pv}.
The bound was generalized in \cite{gpv} by tilting,
via a parameter $\theta \ge 1$, the posterior hypothesis distribution.
The generalized bound was noted to progressively improve
with $\theta$; however its asymptotic formula as $\theta$ tends to infinity
was not determined.

In this paper, we revisit this generalized bound and establish   
its asymptotic expression in closed-form. 
We then investigate the asymptotic generalized bound 
in the classical context of the error probability of block codes used over
the memoryless  binary symmetric channel (BSC) with crossover probability $p< \frac 12$.
We prove that it is exponentially tight for arbitrary sequences of zero-rate codes
and hence achieves the BSC zero-rate error exponent or reliability 
function (for in-depth studies of the channel reliability function, whose 
characterization at low rates remains a long-standing open problem, 
see~\cite{SGB67-I,SGB67-II,gallager,viterbi,
csiszar,blahut,dalai13} and the references therein).

In showing the exponential tightness of
the asymptotic generalized Poor-Verd\'u bound, we first observe
that when a code $\code_n$ with blocklength $n$ and size $|\code_n|=M_n$ 
is transmitted over the BSC, this bound 
exactly equals the probability of the set $\notie(\code_n)$, which consists of 
all input-output $n$-tuple pairs $(x^n,y^n) \in \code_n \times \yc^n$ satisfying
$$d(x^n,y^n)>\min_{u^n\in\code_n\setminus\{x^n\}}d(u^n,y^n),$$
%\quad\forall~x^n\in\code_n,$$
where $d(\cdot,\cdot)$ is the Hamming distance and  $\yc$ is the channel output alphabet 
(see Section~III).
By adding the probability of all ties, i.e., all $(x^n,y^n) \in \code_n \times \yc^n$ such that
$$d(x^n,y^n)=\min_{u^n\in\code_n\setminus\{x^n\}}d(u^n,y^n),$$
%\quad\forall~x^n\in\code_n,$$
which are collected in the set $\tie(\code_n)$,
to $\Pr(\notie(\code_n))$,
an upper bound on 
the minimum probability of decoding error $P_e$ is then obtained.
The exponential tightness of $\Pr(\notie(\code_n))$ to $P_e$ can thus be confirmed by showing that
$\Pr(\tie(\code_n))$ has either the same error exponent as, or decreases exponentially faster than, $\Pr(\notie(\code_n))$ for zero-rate codes.
This property is demonstrated by constructing 
partitions of $\tie(\code_n)$ and 
$\notie(\code_n)$, denoted by $\{\Tc_i\}_{i=1}^{M_n}$ and $\{\Nc_i\}_{i=1}^{M_n}$, respectively,
and then judiciously relating the probability of component set $\Tc_i$ to that of component set
$\Nc_i$ for $i=1,\ldots,M_n$. 
Specifically, we show that 
the probability of a finite cover of each $\Tc_i$,
multiplied by $2M_n\frac{(1-p)}p$, 
is no larger than the probability of a subset of $\Nc_i$
(cf.~Figure~\ref{figure1} in Section~III).
With these key ingredients in place, we prove the exponential tightness 
of the asymptotic generalized Poor-Verd\'u bound at rate zero
(i.e., when $\limsup_{n\rightarrow\infty}\frac 1n\log |\code_n|=0$).

The rest of the paper is organized as follows. In Section~II, the exact expression of the
asymptotic generalized Poor-Verd\'u lower bound on the error probability in multihypothesis 
testing is derived. In Section~III, the error exponent analysis of this asymptotic 
bound is carried out in detail
for the channel coding problem over the BSC. Finally conclusions are drawn in Section~IV.

\section{Asymptotic Expression of the Generalized Poor-Verd\'u Bound}

In 1995, Poor and Verd\'u established a lower bound on the error probability of multihypothesis testing~\cite{pv}. This bound was generalized in~\cite{gpv} in terms of a {\em tilted} posterior hypothesis distribution with tilting parameter $\theta\ge1$ (with the original bound in~\cite{pv}
recovered when $\theta=1$).

\begin{lemma}[Generalized Poor-Verd\'{u} bound \cite{gpv}]\label{newbound}
{\rm
\index{Poor-Verd\'{u} lemma} 
Consider random variables $X$ and $Y$, governed by the joint distribution $P_{X,Y}$, 
and  that take values in a discrete 
(i.e., finite or countably infinite) 
alphabet $\xc$ and an arbitrary alphabet $\yc$, respectively. The minimum
probability of error $P_{\text{e}}$ in estimating $X$ from $Y$ satisfies
\begin{IEEEeqnarray}{rCl}
P_{\text{e}}\geq(1-\alpha)\!\cdot\! P_{X,Y}\bigg\{(x,y)\in\xc\times\yc:
P_{X|Y}^{(\theta)}(x|y)\leq \alpha\bigg\},\quad\,\,\,
\label{gen-pv-bound}
\end{IEEEeqnarray}
for each $\alpha\in[0,1]$ and arbitrary $\theta\geq 1$, where 
\begin{equation*}\label{twistdist}
P_{X|Y}^{(\theta)}(x|y)\bydef \frac{(P_{X|Y}(x|y))^\theta}
{\sum_{u\in\xc}(P_{X|Y}(u|y))^\theta}
\end{equation*}
is the tilted distribution of $P_{X|Y}(x|y)$ with parameter $\theta$.
}\end{lemma}

It is illustrated via examples in \cite{gpv} that the lower bound
in \eqref{gen-pv-bound} improves in general as $\theta$ grows.
However, the asymptotic expression of \eqref{gen-pv-bound},
as $\theta$ goes to infinity, was not established in closed-form.
This issue is resolved in what follows.

\begin{lemma}\label{lemma-infty-nonu} Let distribution $P_X$ have finite support $\code\subseteq\Xc$. Then for $\alpha<\frac 1{|\code|}$, 
\begin{IEEEeqnarray}{rCl}
\lefteqn{\limsup_{\theta\rightarrow\infty}P_{X,Y}\left\{(x,y)\in\xc\times\yc:~P_{X|Y}^{(\theta)}(x|y)\leq \alpha\right\}}\nonumber\\
&=&P_{X,Y}\bigg\{(x,y)\in\xc\times\yc:~ j_{XW}(x;y)
<\max_{u\in\code}j_{XW}(u;y)\bigg\},\nonumber
\end{IEEEeqnarray}
where $W=P_{Y|X}$ typically
denotes a channel transition probability with input $X$ and output $Y$,
and $$j_{XW}(x;y)\triangleq\log P_{X|Y}(x|y).$$
\end{lemma}
\begin{IEEEproof}
Setting $\alpha=e^{-\kappa}$ in the right-hand side (RHS) probability term in~\eqref{gen-pv-bound} yields
\begin{IEEEeqnarray}{rCl}
\lefteqn{P_{X,Y}\left\{(x,y)\in\xc\times\yc:~P_{X|Y}^{(\theta)}(x|y)\leq e^{-\kappa}\right\}}\nonumber\\
&=&P_{X,Y}\left\{(x,y)\in\xc\times\yc:~\frac{e^{\theta\cdot j_{XW}(x;y)}}
{\sum_{u\in\code}e^{\theta\cdot j_{XW}(u;y)}}
\leq e^{-\kappa}\right\}\nonumber\\
&=&P_{X,Y}\bigg\{(x,y)\in\xc\times\yc: \nonumber\\
&&\hspace*{10mm}j_{XW}(x;y)\leq \frac 1{\theta}\log\bigg(\sum_{u\in\code}e^{\theta\cdot j_{XW}(u;y)}\bigg)-\frac {\kappa}{\theta}\bigg\}.\quad\label{inside-nonu}\notag
\end{IEEEeqnarray}

Noting that 
\begin{eqnarray}
\lefteqn{j_{XW}(x;y)\leq \frac 1{\theta}\log\bigg(\sum_{u\in\code}e^{\theta\cdot j_{XW}(u;y)}\bigg)-\frac \kappa{\theta}}\nonumber\\
\hspace{-0.1in} && \hspace{-0.1in} \iff
\frac \kappa{\theta}\leq \frac 1{\theta}\log\bigg(\sum_{u\in\code}e^{\theta\cdot j_{XW}(u;y)}\bigg)-j_{XW}(x;y), \label{con-nonu}
\end{eqnarray}
we separately consider the following two cases.
\begin{itemize}
\item For $(x,y)$ with $j_{XW}(x;y)<\max_{u\in\code}j_{XW}(u;y)$, the RHS of 
\eqref{con-nonu} will approach $$\max_{u\in\code}j_{XW}(u;y)-j_{XW}(x;y)>0$$
as $\theta$ grows without bound, while the left-hand side of \eqref{con-nonu} tends to zero. Hence, \eqref{con-nonu}
holds for $\theta$ sufficiently large. 
\item For $(x,y)$ with $j_{XW}(x;y)=\max_{u\in\code}j_{XW}(u;y)$, 
\begin{IEEEeqnarray}{rCl}
\lefteqn{\frac 1\theta\log\bigg(\sum_{u\in\code}e^{\theta\cdot j_{XW}(u;y)}\bigg)-j_{XW}(x;y)}\nonumber\\
&=&\frac 1\theta\log\bigg(e^{\theta\cdot j_{XW}(x,y)}\sum_{u\in\code}e^{\theta\cdot (j_{XW}(u;y)-j_{XW}(x;y))}\bigg)\nonumber\\
&&\hspace*{50mm}-j_{XW}(x;y)\nonumber\\[2mm]
&=&\frac 1\theta\log\bigg(\sum_{u\in\code}e^{\theta\cdot (j_{XW}(u;y)-j_{XW}(x;y))}\bigg)\nonumber\\
&\leq&\frac 1\theta\log|\code|,\nonumber
\end{IEEEeqnarray}
where the last inequality holds because $j_{XW}(u;y)\leq j_{XW}(x;y)$ for all $u\in\code$. Hence, 
\eqref{con-nonu} is violated since $\kappa=-\log\alpha>\log|\code|$.
\end{itemize}
Verifying the above two cases completes the proof.
\end{IEEEproof}

In light of Lemma~\ref{lemma-infty-nonu}, we can fix $\kappa=-\log\alpha>\log|\code|$,
take $\theta$ to infinity and obtain from \eqref{gen-pv-bound} that
\begin{IEEEeqnarray}{rCl}
\lefteqn{P_{\text{e}}\geq(1-e^{-\kappa})}\nonumber\\
&&\hspace*{2mm} P_{X,Y}\bigg\{(x,y)\in\xc\times\yc:
j_{XW}(x;y)
<\max_{u\in\code}j_{XW}(u;y)\bigg\}.\quad\label{final-bound}
\end{IEEEeqnarray}
Since \eqref{final-bound} holds for $\kappa>\log|\code|$ arbitrarily large,
we have the following asymptotic expression of the generalized Poor-Verd\'{u} bound.
\begin{corollary}\label{corollary1}
The minimum error probability $P_{\text{e}}$ in estimating $X$ from $Y$ satisfies
\begin{IEEEeqnarray}{rCl}
P_{\text{e}}\geq P_{X,Y}\!\bigg\{\!(x,y)\!\in\!\xc\!\times\!\yc\!:\!
j_{XW}(x;y)
\!<\!\max_{u\in\code}j_{XW}(u;y)\!\bigg\}\!.\quad\label{last-bound}
\end{IEEEeqnarray}
\end{corollary}

Two remarks are made based on Corollary~\ref{corollary1}.
First, the optimal estimate of $X$ from observing $Y$
is known to be the maximum \emph{a posteriori} estimate,
given by
\begin{equation}
\label{eq:pv1}
e(y)=\arg \max_{x\in\code}P_{X|Y}(x|y)=\arg \max_{x\in\code}j_{XW}(x;y),
\end{equation}
and the lower bound in \eqref{last-bound} can in fact be deduced directly 
from \eqref{eq:pv1}. This indicates that tilting the \emph{a posteriori} distribution
in the generalized Poor-Verd\'{u} bound can 
indeed approach\footnote{
Note that the set $\big\{(x,y)\in\xc\times\yc:P_{X|Y}(x|y)=P_{X|Y}(e(y)|y)\big\}$
includes all ties. For example, for the 2-fold BSC 
(i.e., the BSC used twice to transmit 2-tuple inputs) with 
uniform $P_X$ over $\code=\{00,11\}$, both $(00,01)$ and $(11,01)$ will be in this set, i.e., 
\begin{IEEEeqnarray*}{rCl}
\lefteqn{\big\{(x,y)\in\xc\times\yc:P_{X|Y}(x|y)=P_{X|Y}(e(y)|y)\big\}}\\
&=&\{(00,00),(00,01),(11,01),(00,10),(11,10),(11,11)\}.
\end{IEEEeqnarray*}
} 
$$1-P_{X,Y}\bigg\{(x,y)\in\xc\times\yc:P_{X|Y}(x|y)=P_{X|Y}(e(y)|y)\bigg\}.$$
As a consequence, the lower bound in \eqref{last-bound}
is tight if and only if 
the $x$ that maximizes $P_{X|Y}(x|y)$ is unique 
for all $y\in\yc$. This elucidates why in the example of \cite[Fig.~1]{gpv}
the generalized Poor-Verd\'u bound achieves the minimum probability of error $P_{\text{e}}$ when $\theta$ grows unbounded.

Second, an alternative lower bound for $P_e$ is the Verd\'{u}-Han bound established in \cite{vh13}. 
This bound was recently generalized in \cite[Thm.~1]{vcfm}. 
We remark that the Verd\'{u}-Han bound is not tight
even if $P_{X|Y}(x|y)$ admits a unique maximizer for every $y \in \yc$. 
For example,  we can obtain
from the ternary hypothesis testing example
in \cite[Sec.~III-A]{gpv} and \cite[Sec.~III-A]{vcfm} that:
\begin{IEEEeqnarray}{rCl}
P_e=\frac 35&>&\max_{\gamma\geq 0}\bigg(\Pr\left[P_{X|Y}(X|Y)\leq\gamma\right]-\gamma\bigg)=\frac{27}{47},\label{vh-bound}
\end{IEEEeqnarray}
where the maximizer in \eqref{vh-bound} is $\gamma^*=\frac{20}{47}$. 
Noting the sub-optimality of the Verd\'u-Han bound, the authors in \cite{vcfm} 
generalized it by varying the output statistics. They also proved the tightness of the resulting generalized Verd\'{u}-Han bound:
\begin{IEEEeqnarray}{rCl}
P_e=\max_{Q_Y}\max_{\gamma\geq 0}\bigg(\Pr\left[\frac{P_{X,Y}(X,Y)}{Q_Y(Y)}\leq\gamma\right]-\gamma\bigg).\label{gvh-bound}
\end{IEEEeqnarray}
It is pertinent to note that the maximizers of~\eqref{gvh-bound}
are given by 
$$\gamma^\ast=\int_{\yc}\max_{x\in\xc}P_{X,Y}(x,y)\,{\text d}P_Y(y)=1-P_e$$ and 
\begin{IEEEeqnarray}{rCl}
Q_Y^\ast(y)&=&\frac{\max_{x\in\xc}P_{X,Y}(x,y)}{
\int_{\yc}\max_{x\in\xc}P_{X,Y}(x,y)\,{\text d}P_Y(y)}\label{op-q}\\
&=&\frac{P_Y(y)P_{X|Y}(e(y)|y)}{1-P_e}.\notag
\end{IEEEeqnarray}
%where we derive
%\begin{IEEEeqnarray}{rCl}
%\lefteqn{\Pr\left[\frac{P_{X,Y}(X,Y)}{Q_Y^\ast(Y)}\leq\gamma^\ast\right]-\gamma^\ast}\notag\\
%&=&P_{X,Y}\big\{(x,y)\in\xc\times\Yc:\nonumber\\
%&&~~~~P_{X|Y}(x|y)\leq \max_{u\in\code}P_{X|Y}(u|y)\big\}-(1-P_e)\notag\\
%&=&P_e.\notag
%\end{IEEEeqnarray}
Hence, the determination of the maximizers of the above generalized Verd\'u-Han bound
is equivalent to determining the minimum error probability $P_e$ itself.

Similar to the generalized Poor-Verd\'u bound with parameter $\theta$, any $Q_Y$ and $\gamma$ adopted for the generalized Verd\'u-Han bound 
yields a lower bound on $P_e$. However, an interesting difference between the generalized Poor-Verd\'u bound and the generalized 
Verd\'u-Han bound is that 
when $P_X$ is uniformly distributed over its support $\code$,
the former bound can be transformed into a function of the \emph{information density} 
$$i_{XW}(x,y)\triangleq\frac{P_{Y|X}(y|x)}{P_Y(y)},$$
while the latter bound cannot.
This transformation may facilitate the interpretation  
of the error exponent via the information density (or equivalently, the Hamming distance) 
for memoryless symmetric channels such as the BSC.

\section {Exponential Tightness of the Asymptotic Generalized Poor-Verd\'u Bound for the BSC at Zero Rate}

In this section, we prove that the asymptotic expression of the generalized Poor-Verd\'u bound
given in \eqref{last-bound} exactly characterizes the zero-rate coding error exponent of the BSC
with crossover probability $p< \frac 12$.
Note that while the error exponent formula for the BSC at zero-rate, $E(0)$, is already known,
$E(0) = -\frac{1}{4} \ln \big(4p(1-p)\big)$ \cite{gallager}, we do not explicitly calculate it.
Rather, we demonstrate that the bound in \eqref{last-bound} is exponentially tight for arbitrary
sequences of zero-rate block codes used over the BSC, hence indirectly achieving $E(0)$. This
approach may be beneficial for a larger class of channels.

Fix a sequence of codes $\{\code_n\}_{n=1}^\infty$ of blocklength $n$, with $\code_n \subseteq \{0,1\}^n$, 
and let $P_{X^n}$ be the uniform distribution over $\code_n$, where $X^n$ denotes the $n$-tuple $(X_1,\ldots,X_n)$.
Denote by $$a_n\triangleq P_{\text{e}}(\code_n)$$ 
the minimum probability of decoding error for transmitting code $\code_n$ over the BSC with crossover probability $p< \frac 12$, and let $b_n$ denote the RHS of~\eqref{last-bound} in this channel coding context:
\begin{IEEEeqnarray}{rCl}
b_n&\triangleq&P_{X^n,Y^n}\!\bigg\{\!(x^n,y^n)\!\in\!\xc^n\!\times\!\yc^n\!:\nonumber\\
&&~~~~~
j_{X^nW^n}(x^n;y^n)
\!<\!\max_{u^n\in\code_n}j_{X^nW^n}(u^n;y^n)\!\bigg\}\nonumber\\
&=&{P_{X^n,Y^n}\bigg\{(x^n,y^n)\in\xc^n\times\yc^n:}\nonumber\\
&&~~~~~
P_{X^n|Y^n}(x^n|y^n)<\max_{u^n\in\code_n}P_{X^n|Y^n}(u^n|y^n)\bigg\}.\label{con}
\end{IEEEeqnarray}
Since the BSC has $p< \frac 12$, the inequality condition in \eqref{con} can be equivalently characterized via
the Hamming distance $d(\cdot,\cdot)$. Hence, 
$$b_n=P_{X^n,Y^n}(\notie(\code_n)),$$
where
\begin{IEEEeqnarray}{rCl}
\lefteqn{\notie(\code_n)\triangleq\bigg\{(x^n,y^n)\in\code_n\times\yc^n:}\nonumber\\
&&~~~~~~~~~~~~d(x^n,y^n)>\min_{u^n\in\code_n\setminus\{x^n\}}d(u^n,y^n)\bigg\}.\nonumber
\end{IEEEeqnarray}
Define the \emph{set of ties} with respect to code $\code_n$ as 
\begin{IEEEeqnarray}{rCl}
\lefteqn{\tie(\code_n)\triangleq\bigg\{(x^n,y^n)\in\code_n\times\yc^n:}\nonumber\\
&&~~~~~~~~~~~~d(x^n,y^n)=\min_{u^n\in\code_n\setminus\{x^n\}}d(u^n,y^n)\bigg\},\nonumber
\end{IEEEeqnarray}
and let 
$$\delta_n=P_{X^n,Y^n}(\tie(\code_n)).$$
Then,
\begin{equation}
b_n\leq a_n\leq b_n+\delta_n,\label{bound}
\end{equation}
which implies that 
\begin{equation}
0\leq \frac 1n\log\frac{a_n}{b_n}\leq \frac 1n\log\left(1+\frac{\delta_n}{b_n}\right).\nonumber
\end{equation}
As a result, 
in order to prove that $a_n$ and $b_n$ have the same error exponent,
it suffices to prove that
\color{black}
\begin{equation}
\limsup_{n\rightarrow\infty}\frac 1n\log\left(1+\frac{\delta_n}{b_n}\right)=0.\label{zero}
\end{equation}
We next establish the following main theorem, which confirms \eqref{zero} at zero rates (in 
Corollary~\ref{corollary2} below).

\begin{theorem}\label{theorem1} 
For any sequence of codes $\{\code_n\}_{n=1}^\infty$, we have
\begin{equation}
\limsup_{n\rightarrow\infty}\frac 1n\log\frac{a_n}{b_n}\leq \limsup_{n\rightarrow\infty}\frac 1n\log M_n,\label{thm1-eq}
\end{equation}
where $M_n=|\code_n|$.
\end{theorem}

\begin{figure}[t]
\begin{center}
    \includegraphics[width=0.8\linewidth]{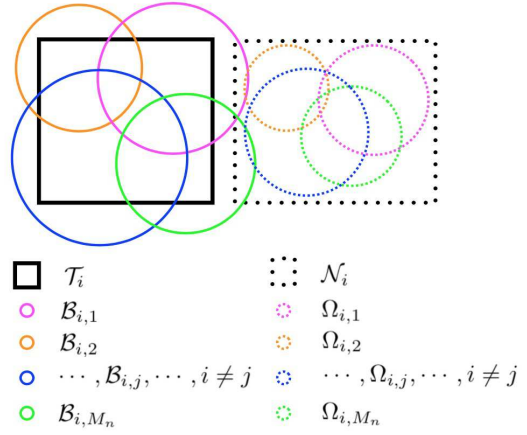}\\
\caption{Illustration of the idea behind the proof of Theorem~\ref{theorem1}.}\label{figure1}
\end{center}
\end{figure}

Before giving the proof, we elucidate the underlying idea behind it.
We first introduce the following necessary notation.
For block code $\code_n=\{x_{(1)}^n,x_{(2)}^n,\ldots,x_{(M_n)}^n\}$ consisting of
$M_n$ distinct codewords, define the sets
\begin{equation}
\Tc_i\triangleq\{y^n\in\yc^n: (x^n_{(i)},y^n)\in\tie(\code_n)\} \nonumber
\end{equation}
for $i=1,\ldots,M_n$. We can then write
\begin{IEEEeqnarray}{rCl}
\delta_n&=&P_{X^n,Y^n}(\tie(\code_n))\nonumber\\
%&=&P_{X^n,Y^n}( \cup_{i=1}^{M_n}\{x^n_{(i)}\times \Tc_i\})\nonumber\\
%&=&\sum_{i=1}^{M_n}P_{X^n,Y^n}(x^n_{(i)}\times \Tc_i)\nonumber\\
&=&\sum_{i=1}^{M_n} P_{X^n}(x_{(i)}^n)\Pr\left(Y^n\in \Tc_i\left|X^n=x_{(i)}^n\right.\right).\label{del}
\end{IEEEeqnarray}
Similarly, defining the sets
\begin{equation}
\Nc_i\triangleq\{y^n\in\yc^n: (x^n_{(i)},y^n)\in\notie(\code_n)\} \nonumber
\end{equation}
for $i=1,\ldots,M_n$, we have
\begin{IEEEeqnarray}{rCl}
b_n&=&P_{X^n,Y^n}(\notie(\code_n))\nonumber\\
%&=&P_{X^n,Y^n}( \cup_{i=1}^{M_n}\{x^n_{(i)}\times \Nc_i\})\nonumber\\
%&=&\sum_{i=1}^{M_n}P_{X^n,Y^n}(x^n_{(i)}\times \Nc_i)\nonumber\\
&=&\sum_{i=1}^{M_n} P_{X^n}(x_{(i)}^n)\Pr\big(Y^n\in\Nc_i\left|X^n=x_{(i)}^n\right.\big).\label{bbb}
\end{IEEEeqnarray}
Finally for $i,j=1,\ldots,M_n$ with $i \neq j$, define
\begin{IEEEeqnarray}{rCl}
\Bc_{i,j}&\triangleq&\left\{y^n\in\yc^n:d(x_{(i)}^n,y^n)=d(x_{(j)}^n,y^n)\right\},\nonumber
\end{IEEEeqnarray}
and
\begin{IEEEeqnarray}{rCl}
\Omega_{i,j}\triangleq\left\{y^n\in\yc^n:d(x_{(i)}^n,y^n)>d(x_{(j)}^n,y^n)\right\}.\notag
\end{IEEEeqnarray}
Then as shown in Fig.~\ref{figure1}, 
we have that $\cup_{j=1,j\neq i}^{M_n} \Bc_{i,j}$ is a finite cover of $\Tc_i$, i.e., 
$$\Tc_i\subseteq \cup_{j=1,j\neq i}^{M_n} \Bc_{i,j}.$$ Hence,
\begin{IEEEeqnarray}{rCl}
\lefteqn{\Pr\big(Y^n\in \Tc_i|X^n=x^n_{(i)}\big)}\notag\\
&\leq&
\Pr\left(\left.Y^n\in\bigcup_{j=1,j\neq i}^{M_n} \Bc_{i,j}\right|X^n=x_{(i)}^n\right)\notag\\
&\leq&\sum_{j=1,j\neq i}^{M_n}\Pr\big(Y^n\in \Bc_{i,j}|X^n=x^n_{(i)}\big)\notag\\
&\leq&(M_n-1)\max_{1\leq j\leq M_n,j\neq i}
\hspace{-0.05in} \Pr\big(Y^n\in \Bc_{i,j}|X^n=x^n_{(i)}\big)\label{t-max}\\
&=&(M_n-1)\,\Pr\big(Y^n\in \Bc_{i,j_i^\ast}|X^n=x^n_{(i)}\big),\notag
\end{IEEEeqnarray}
where the second inequality follows from the union bound and
$j_i^\ast$ is the maximizer of \eqref{t-max}. 
Next, noting that
$$\Omega_{i,j}\subseteq\Nc_i$$ for all $1\leq j\leq M_n$ and $j\neq i$, we have
\begin{IEEEeqnarray*}{rCl}
\Pr\big(Y^n\in \Omega_{i,j_i^\ast}|X^n=x^n_{(i)}\big)
\leq \Pr\big(Y^n\in \Nc_i|X^n=x^n_{(i)}\big).
\end{IEEEeqnarray*}
Thus, if $\Pr\big(Y^n\in \Bc_{i,j_i^\ast}|X^n=x^n_{(i)}\big)$ and $\Pr\big(Y^n\in \Omega_{i,j_i^\ast}|X^n=x^n_{(i)}\big)$ are of comparable order in the sense that
\begin{IEEEeqnarray}{rCl}
\lefteqn{\Pr\left(Y^n\in \Omega_{i,j_i^\ast}\left|X^n=x_{(i)}^n\right.\right)}\notag\\
&\geq&c\cdot \Pr\left(Y^n\in \Bc_{i,j_i^\ast}\left|X^n=x_{(i)}^n\right.\right)\notag
\end{IEEEeqnarray}
for some constant $c$ independent of $n$ 
and $i$, 
then we have
\begin{IEEEeqnarray}{rCl}
\Pr\left(Y^n\in \Nc_i\left|X^n=x_{(i)}^n\right.\right)
&\geq&\frac{c}{M_n}\Pr\left(Y^n\in \Tc_i\left|X^n=x_{(i)}^n\right.\right),\notag
\end{IEEEeqnarray}
which immediately gives
$$b_n\geq\frac c{M_n}\delta_n$$ and confirms \eqref{thm1-eq}.
With this idea in mind, we next provide the detailed proof.

\begin{IEEEproof}[Proof of Theorem \ref{theorem1}]
%Let $\ML(\code_n)$ denote the set of $(x^n,y^n)$ pairs that result in an erroneous decision for a non-randomized maximum-likelihood (ML) decoder. 
\begin{enumerate}
\item First, we calculate $\Pr\big(Y^n\in \Bc_{i,j}\big|x_{(i)}^n\big)$.

\hspace*{4mm}
For each $x_{(i)}^n$ and $x_{(j)}^n$, if $d(x_{(i)}^n,x_{(j)}^n)=2\ell\geq 2$ is even, then there are $\binom{2\ell}\ell\binom{n-2\ell}{m}$ of 
$y^n$'s such that $d(x_{(i)}^n,y^n)=d(x_{(j)}^n,y^n)=\ell+m$ for $0\leq m\leq n-2\ell$; else if $d(x_{(i)}^n,x_{(j)}^n)=2\ell-1$ is odd, then there exist no $y^n$ such that
$d(x_{(i)}^n,y^n)=d(x_{(j)}^n,y^n)$. 
As a result, we have that
\begin{IEEEeqnarray}{rCl}
\lefteqn{\Pr\left(Y^n\in \Bc_{i,j}\left|x_{(i)}^n\right.\right)}\notag\\
&=&\begin{cases}
\displaystyle\sum_{m=0}^{n-2\ell}\mbox{$\binom{2\ell}\ell$}\mbox{$\binom{n-2\ell}{m}$}
(1-p)^{n-\ell-m}p^{\ell+m},\\
\hspace*{35mm}\text{if }d(x_{(i)}^n,x_{(j)}^n)=2\ell;\\
0,\hspace*{32mm}\text{if }d(x_{(i)}^n,x_{(j)}^n)=2\ell-1\\
\end{cases}\nonumber\\
&=&\begin{cases}
\displaystyle\mbox{$\binom{2\ell}\ell$} p^\ell (1-p)^{\ell},&\text{if }d(x_{(i)}^n,x_{(j)}^n)=2\ell;\\
0,&\text{if }d(x_{(i)}^n,x_{(j)}^n)=2\ell-1.\\
\end{cases}\label{184}
\end{IEEEeqnarray}

\item We next lower-bound $\Pr\big(Y^n\in \Omega_{i,j}\big|x_{(i)}^n\big)$
in terms of $\Pr\big(Y^n\in \Bc_{i,j}\big|x_{(i)}^n\big)$.

\hspace*{4mm} If $d(x_{(i)}^n,x_{(j)}^n)=2\ell$ is even, there are 
\begin{equation}
\sum_{\ell'=0}^{\min\{m,\ell-1\}}\mbox{$\binom{2\ell}{\ell+\ell'+1}$}\mbox{$\binom{n-2\ell}{m-\ell'}$}
\label{count1}
\end{equation}
of $y^n$'s
satisfying $d(x_{(i)}^n,y^n)=\ell+1+m$ and $d(x_{(i)}^n,y^n)>d(x_{(j)}^n,y^n)$
for $0\leq m\leq n-2\ell$; else if $d(x_{(i)}^n,x_{(j)}^n)=2\ell-1$ is odd, then 
there are 
\begin{equation}\sum_{\ell'=0}^{\min\{m,\ell\}}\mbox{$\binom{2\ell-1}{\ell+\ell'+1}$}\mbox{$\binom{n-2\ell+1}{m-\ell'}$}\label{count2}
\end{equation} of $y^n$'s
satisfying $d(x_{(i)}^n,y^n)=\ell+1+m$ and $d(x_{(i)}^n,y^n)>d(x_{(j)}^n,y^n)$
for $0\leq m\leq n-2\ell+1$. 

\hspace*{4mm}
Taking $\ell'=0$ in \eqref{count1} and \eqref{count2} gives a lower bound 
on $\Pr\big(Y^n\in \Omega_{i,j}\big|x_{(i)}^n\big)$ as follows:
\begin{IEEEeqnarray}{rCl}
\lefteqn{\Pr\left(Y^n\in \Omega_{i,j}\left|X^n=x_{(i)}^n\right.\right)}\notag\\
%&&=\begin{cases}
% \displaystyle\sum_{m=0}^{n-2\ell}\sum_{\ell'=0}^{\min\{m,\ell-1\}}\binom{2\ell}{\ell+\ell'+1}\binom{n-2\ell}{m-\ell'} (1-p)^{n-\ell-1-m}p^{\ell+1+m},&d(x_{(i)}^n,x_{(j_i^\ast)}^n)\text{ even}\\
% \displaystyle\sum_{m=0}^{n-2\ell+1}\sum_{\ell'=0}^{\min\{m,\ell\}}\binom{2\ell-1}{\ell+\ell'+1}\binom{n-2\ell+1}{m-\ell'}(1-p)^{n-\ell-1-m}p^{\ell+1+m},&d(x_{(i)}^n,x_{(j_i^\ast)}^n)\text{ odd}\\
%\end{cases}\notag\\
&\geq&\begin{cases}
\displaystyle\sum_{m=0}^{n-2\ell}\mbox{$\binom{2\ell}{\ell+1}$}\mbox{$\binom{n-2\ell}{m}$} (1-p)^{n-\ell-1-m}p^{\ell+1+m},\\
\hspace*{35mm}\text{ if }d(x_{(i)}^n,x_{(j)}^n)=2\ell\\
 \displaystyle\sum_{m=0}^{n-2\ell+1}\mbox{$\binom{2\ell-1}{\ell+1}$}\mbox{$\binom{n-2\ell+1}{m}$}(1-p)^{n-\ell-1-m}p^{\ell+1+m},\\
 \hspace*{35mm}\text{ if }d(x_{(i)}^n,x_{(j)}^n)=2\ell-1
\end{cases}\notag\\
&=&\begin{cases}
\displaystyle\mbox{$\binom{2\ell}{\ell+1}$}p^{\ell+1}(1-p)^{\ell-1},&\text{ if }d(x_{(i)}^n,x_{(j)}^n)=2\ell\\
 \displaystyle\mbox{$\binom{2\ell-1}{\ell+1}$}p^{\ell+1}(1-p)^{\ell-2},&\text{ if }d(x_{(i)}^n,x_{(j)}^n)=2\ell-1
\end{cases}\notag\\
&\geq&
\frac{\ell}{(\ell+1)}\frac{p}{(1-p)}\Pr\left(Y^n\in \Bc_{i,j}\left|X^n=x_{(i)}^n\right.\right)\label{186}\\
&\geq&\frac{p}{2(1-p)}\Pr\left(Y^n\in \Bc_{i,j}\left|X^n=x_{(i)}^n\right.\right),\label{187a}
\end{IEEEeqnarray}
where \eqref{186} follows from \eqref{184} and \eqref{187a} holds since $\ell\geq 1$.

\item We next can write
\begin{IEEEeqnarray}{rCl}
\Tc_i&=&\left\{y^n\in\yc^n:d(x_{(i)}^n,y^n)=\min_{u^n\in\code_n\setminus\{x_{(i)}^n\}}d(u^n,y^n)\right\}\notag\\
&\subseteq&\bigcup_{j=1,j\neq i}^{M_n} \left\{y^n\in\yc^n:d(x_{(i)}^n,y^n)=d(x_{(j)}^n,y^n)\right\}\notag\\
&=&\bigcup_{j=1,j\neq i}^{M_n} \Bc_{i,j},\notag
\end{IEEEeqnarray}
which implies, as already shown in~\eqref{t-max}, that  
\begin{IEEEeqnarray}{rCl}
\lefteqn{\Pr\left(Y^n\in\Tc_i\left|X^n=x_{(i)}^n\right.\right)}\notag\\
%&\leq&
%\Pr\left(\left.Y^n\in\bigcup_{j=1,j\neq i}^{M_n} \Bc_{i,j}\right|X^n=x_{(i)}^n\right)\notag\\
%&\leq&\sum_{j=1,j\neq i}^{M_n} \Pr\left(Y^n\in \Bc_{i,j}\left|X^n=x_{(i)}^n\right.\right)\label{union}\\
&\leq&(M_n-1)\Pr\left(Y^n\in \Bc_{i,j_i^\ast}\left|X^n=x_{(i)}^n\right.\right),
\label{182}
\end{IEEEeqnarray}
where $j_i^\ast$ is the maximizer in~\eqref{t-max}.  
%\begin{equation}
%j_i^\ast\triangleq\arg\max_{1\leq j\leq M_n} \Pr\left(Y^n\in \Bc_{i,j}\left|X^n=x_{(i)}^n\right.\right).\label{183}
%\end{equation}
Therefore with this $j_i^\ast$, 
%as defined in \eqref{183}, 
we have that
\begin{IEEEeqnarray}{rCl}
\lefteqn{\Pr\left(Y^n\in \Nc_i\left|X^n=x_{(i)}^n\right.\right)}\notag\\
&\geq&\Pr\left(Y^n\in \Omega_{i,j_i^\ast}\left|X^n=x_{(i)}^n\right.\right)\notag\\
&\geq&\frac{p}{2(1-p)}\Pr\left(Y^n\in \Bc_{i,j_i^\ast}\left|X^n=x_{(i)}^n\right.\right)\label{187}\\
&\geq&\frac{p}{2(1-p)}\frac{1}{(M_n-1)}\Pr\left(Y^n\in \Tc_i\left|X^n=x_{(i)}^n\right.\right),\quad\quad\label{188}
\end{IEEEeqnarray}
where
\eqref{187} follows from \eqref{187a},
and \eqref{188} is based on \eqref{182}.

\item We conclude from \eqref{188} that
\begin{IEEEeqnarray}{rCl}
b_n
&=&\sum_{i=1}^{M_n} P_{X^n}(x_{(i)}^n)\Pr\left(Y^n\in\Nc_i\left|X^n=x_{(i)}^n\right.\right)\notag\\
&\geq&\frac{p}{2(1-p)}\frac{1}{(M_n-1)}\cdot\nonumber\\
&&\sum_{i=1}^{M_n} P_{X^n}(x_{(i)}^n)\Pr\left(Y^n\in\Tc_i\left|X^n=x_{(i)}^n\right.\right)\notag\\
&=&\frac{p}{2(1-p)}\frac{1}{(M_n-1)}\delta_n,\label{bd}
\end{IEEEeqnarray}
which implies that
\begin{IEEEeqnarray}{rCl}
 \lefteqn{\limsup_{n\rightarrow\infty}\frac 1n\log\left(1+\frac{\delta_n}{b_n}\right)}\nonumber\\&\leq&
\limsup_{n\rightarrow\infty}\frac 1n\log\left(1+\frac{2(1-p)}{p}(M_n-1)\right)\notag\\
&=&\limsup_{n\rightarrow\infty}\frac 1n\log(M_n),\notag
\end{IEEEeqnarray}
where the last step holds whether either $M_n$ is bounded or unbounded.
\end{enumerate}
\end{IEEEproof}

Finally, we directly obtain that \eqref{zero} holds 
when the (asymptotic) rate of the code sequence 
considered in Theorem~\ref{theorem1} is zero, hence 
confirming the exponential tightness of the asymptotic 
generalized Poor-Verd\'{u} bound for the BSC at rate zero.

\begin{corollary}\label{corollary2}
For any sequence of zero-rate codes $\{\code_n\}_{n=1}^\infty$ used over the BSC, we have
\begin{equation*}
\limsup_{n\rightarrow\infty}\frac 1n\log\frac{a_n}{b_n}\leq \limsup_{n\rightarrow\infty}\frac 1n\log |\code_n|=0.
\end{equation*}
\end{corollary}

\begin{remark}\label{rem1}
It is worth emphasizing that Corollary~\ref{corollary2} does {\it not} hold for the memoryless
binary erasure channel (BEC); i.e., the asymptotic generalized Poor-Verd\'{u} bound is not
exponentially tight for this channel. Indeed for the BEC, the bound in~\eqref{gen-pv-bound}  
is unchanged for every $\theta \ge 1$ (including when $\theta \to \infty$) and is hence 
identical to the original Poor-Verd\'{u} bound. The latter bound was shown in~\cite{vio} 
not to achieve the BEC's error exponent at low rates.
\end{remark}

\section{Conclusion}
We derived a closed-form formula for the asymptotic generalized Poor-Verd\'{u} error bound
to the multihypothesis testing error probability and proved that, 
unlike the case for the BEC~\cite{vio}, it
achieves the zero-rate error coding exponent of the BSC.  

In the proof of Theorem~\ref{theorem1}, 
we used the union bound in the derivation of \eqref{182}, 
%and \eqref{182}, 
which may be loose  when the sequence of codes is no longer of zero rate.
Thus, if a sharper bound can be employed, 
the multiplicative factor $\frac p{2(1-p)}\frac 1{(M_n-1)}$ in \eqref{bd} 
may be improved. We conjecture that Corollary \ref{corollary2}
holds not just for zero-rate codes but that it 
can be indeed extended to arbitrary code sequences 
of positive rate. Proving this conjecture  is an interesting future direction. 
Other future work includes the further examination of tight bounds for codes with small
blocklength (e.g., see~\cite{polyanskiy,CLM13,vcfm}) used over channels with and without memory.

\end{document}